# Investigation of Band-Offsets at Monolayer-Multilayer MoS₂ Junctions by Scanning Photocurrent Microscopy

Sarah L. Howell[§,1,4] Deep Jariwala[§,1,4] Chung-Chiang Wu,[1,4] Kan-Sheng Chen,[1] Vinod K. Sangwan,[1,4] Junmo Kang,[1,4] Tobin J. Marks,[1,2,4] Mark C. Hersam,[1,2,3,4] and Lincoln J. Lauhon*[1,4]

[1]Department of Materials Science and Engineering, Northwestern University, Evanston, IL 60208, USA.

[2]Department of Chemistry, Northwestern University, Evanston, IL 60208, USA.

[3]Department of Medicine, Northwestern University, Evanston, IL 60208, USA.

[4]Materials Research Center, Northwestern University, Evanston, IL 60208, USA.

[§]These authors contributed equally.

*Corresponding author: E-mail: lauhon@northwestern.edu

Abstract:

The thickness-dependent band structure of MoS₂ implies that discontinuities in energy bands exist at the interface of monolayer (1L) and multilayer (ML) thin films. The characteristics of such heterojunctions are analyzed here using current versus voltage measurements, scanning photocurrent microscopy, and finite element simulations of charge carrier transport. Rectifying I-V curves are consistently observed between contacts on opposite sides of 1L-ML junctions, and a strong bias-dependent photocurrent is observed at the junction. Finite element device simulations with varying carrier concentrations and electron affinities show that a type II band alignment at single layer/multi-layer junctions reproduces both the rectifying electrical characteristics and the photocurrent response under bias. However, the zero-bias junction photocurrent and its energy dependence are not explained by conventional photovoltaic and photothermoelectric mechanisms, indicating the contributions of hot carriers.

Transition metal dichalcogenides (TMDCs) such as $MoS_2$ consist of discrete two-dimensional (2D) layers bound together by van der Waals forces, with important consequences for both physical and electronic structure of these ultrathin semiconducting crystals[2,3]. First, "flakes" can be exfoliated by the "Scotch tape" method[4], and such flakes exhibit distinctive thickness-dependent variations in physical properties that are readily detected by optical microscopy[5]. Second, the band structure varies with multilayer thickness, transitioning from a direct bandgap in single layer (1L) $MoS_2$ to an indirect bandgap in two-layer (2L) and multilayer materials[6, 7]. The thickness-dependent band structure implies the existence of discontinuities in energy bands, i.e., heterojunctions, arising from abrupt discontinuities in physical thickness. Such geometry-induced heterojunctions provide an intriguing degree of freedom to engineer optoelectronic devices such as photodetectors [1, 8-12] and energy conversion devices[13-17], but with new operating principles that are distinct from those of more "conventional" lateral[18-21] and vertical 2-D heterojunctions[22-24]. For example, the photocurrent responses at monolayer steps in graphene[25] and topological insulators[26] were explained in terms of changes in the Seebeck coefficient with layer thickness. While one might expect photovoltaic and photothermal effects at thickness discontinuities in $MoS_2$, the band alignments in 1L, 2L, and multilayer films have not been definitively established; reported values for the electron affinity of $MoS_2$ range from 3.74 to 4.45[27-32], reflecting the challenges in calculating the absolute positions of energy bands using first principles computations. Experimentally, it is also difficult to determine the change in electron affinity across a 1L-2L junction in a thin-film using ultraviolet photoelectron spectroscopy measurements, which generally sample large areas. Knowledge of electronic band alignments that govern transport through heterojunctions is a critical foundation of device engineering, yet there is a clear knowledge gap for 2-D materials.

Here we report investigations of the electrical properties and photoresponse of junctions between monolayer and multilayer $MoS_2$ that directly probe the influence of band alignment on heterojunction properties. We find that 1L-ML junctions are rectifying in a consistent direction and exhibit a local photocurrent response that is distinct from that of monolayer and multilayer regions. Finite element device simulations of both the *I-V* characteristics and the photoresponse are conducted to explore the influence of carrier concentration and electron affinity on rectification and photocurrent generation. We find that a type II band alignment at single layer/multi-layer junctions reproduces both the rectifying electrical characteristics and the photocurrent response under bias, and we conclude that the photovoltaic response dominates the photothermoelectric response. However, the zero-bias junction photocurrent does not arise from conventional photovoltaic or photothermoelectric mechanisms. Rather, the energy dependence indicates contributions from hot carriers. Hence, the combined experimental and modeling study addresses a key knowledge gap in 2-D band alignments and reveals novel photocurrent generation mechanisms that can occur at geometrical heterojunctions.

MoS$_2$ flakes were exfoliated from commercially available crystals of molybdenite onto n$^+$ Si substrates coated with 300 nm of SiO$_2$[33], and then optical microscopy and Raman spectroscopy at 532 nm were used to identify junctions between 1L and multilayer (≥2L) regions for subsequent device fabrication (Figs. 1a, b inset). The exfoliation of natural crystals implies that the crystallographic orientation between the monolayer and multilayer regions is consistent between devices, which is important since a twist angle between stacked flakes would affect the band structure[34]. Devices were formed by electron beam lithography and lift-off of Au ohmic contacts (75 nm thick) on opposite sides of junctions producing field effect transistors (FETs) with discrete variations in channel thickness (Fig. 1a), where the drain (source) contact was on the thicker (thinner) MoS$_2$ side. The 1L/2L device of Figure 1a exhibits nonlinear, rectifying behavior in accumulation (Fig. 1b) that is not observed for ohmically contacted devices of uniform thickness. Defining the rectification ratio as the ratio of the magnitude of the current at negative $V_D$ to the current at positive $V_D$, Fig. 1c plots rectification ratios for 6 devices; values of up to 25 are consistently observed, with ratios greater than 1 observed at all $V_g$ when the ML contact is biased in 5 out of 6 devices. Hence, the rectification does not arise from random variations in contact resistance. Although gate tunability has recently been demonstrated to be a useful property of ultrathin junctions[9, 17, 35], the present rectification ratios do not exhibit any obvious trend with multilayer thickness or gate bias in accumulation (Fig. 1c).

To explore the origin of the junction characteristics, detailed electrical and opto-electrical studies were carried out on devices both with uniform channel thickness and with 1L/ML junctions. Our experimental strategy was to establish device modeling parameters based on analysis of uniform thickness (1L, 2L, etc.) thin-film transistors, and then examine whether these parameters could be used to describe the electrical characteristics of the junctions. In addition to conventional current versus voltage (*I-V*) measurements, scanning photocurrent microscopy (SPCM) was used to profile energy bands in operating devices[1] under ambient conditions. We have previously used SPCM to identify the source of rectification in p-n junctions fabricated from p-type single walled carbon nanotubes and $MoS_2$,[9] and SPCM has also been employed to explore the electronic properties of monolayer steps in graphene and bismuth/antimony telluride.[25, 26] The present experimental studies were complemented with finite element modeling using a commercial device simulator (Sentaurus TCAD) for solving the Poisson and continuity equations. Initial materials parameters for the simulations (Supp. Table S1) were established by fitting the output and transfer characteristics of 1L and 4L transistors (Supp. Fig. S1). Similar devices exhibit band-like transport at room temperature.[1, 33] Materials parameters were further tuned by fitting to SPCM profiles as described below.

To generate SPCM images, the sample is scanned beneath a laser beam while simultaneously recording the photocurrent (Fig. 2a), which may arise from gradients in potential (drift) and carrier concentration (diffusion). The reflected light is recorded simultaneously to enable alignment of the photocurrent map with the device structure, and one-dimensional SPCM profiles (Fig. 2c) are extracted from images (Fig. 2b). Illumination was provided by a coherent white light source (NKT Photonics) that was tuned to a specific wavelength, focused by a 100×/0.9NA objective on the device, and modulated at 1837 Hz, which is much lower than the inverse of the device response time (>30 kHz) limited by the preamplifier bandwidth (Supp. Fig. S2)). The power was maintained at 40 μW, and the photoinduced current modulation at 1873 Hz was measured with a current preamplifier and a lock-in amplifier. The lock-in response is proportional to the steady-state device response, but with improved signal-to-noise ratio.

Fig. 2 presents experimental and simulated SPCM profiles for different drain biases. The photoresponse near the contacts at 0 $V_D$ (Fig. 2b,c) is due to both photothermoelectric effects[36] and band bending (i.e., a photovoltaic response)[1].  A finite bias generates an electric field in the channel and leads to changes in band bending near the contacts (Fig. 2d). 2D simulations of free carrier transport following photoexcitation (Fig. 2c) reproduce the experimental SPCM profiles when a low quantum yield is used to account for the large excitonic binding energy,[37] which limits the number of *free* carriers created per absorbed photon. For simplicity, we assume a spatially uniform quantum yield, while noting that additional excitons dissociate into free carriers when the electric field increases near junctions. Because the variation in photocurrent under different applied biases is well-described by simulations of free carrier transport (Fig. 2c), we conclude that the response near the contacts is dominated by photogenerated carrier separation due to band bending as we proposed in earlier work.[1] Contributions from photothermal effects are discussed further below. We also note that trions, or charged excitons, will also form under conditions of electron accumulation[38], but should readily dissociate at room temperature in the presence of moderate electric fields.

The simulated photocurrent profiles for biased devices are sensitive to the choice of free carrier concentration (Fig. 2e). At increased carrier concentrations, the resistance of the contacts and space-charge regions decreases. A larger photocurrent in the channel is observed because the electric field in the channel is larger and the overall series resistance is lower. In contrast, the relative potential drop at the contacts increases with decreasing carrier concentration, so a larger response is seen at the positively biased contact than in the channel. Variations in electron affinity $\chi$ have similar effects (Supp. Fig. S3); as the electron affinity is decreased, charge transfer from the semiconductor to the metal increases, increasing the band bending near the contacts and producing a greater variation in photoresponse under bias. The SPCM response to varying gate voltage is consistent with the trends described above, but the signals at the contacts does not change sign within the range of gate voltages examined.

The bias-dependent photoresponse of a $MoS_2$ single layer/three layer junction device is shown in Fig. 3. Large bias dependent photocurrents are observed at both the contacts and the 1L/3L junction. The photoresponse of the device was simulated (Fig. 3c) using carrier concentrations and electron affinities derived from simulations of uniform channel thickness devices as described in the previous paragraph. In both the experiments (Fig. 3b) and simulations (Fig. 3c), the photoresponse of the 1L/3L junction varies more strongly with drain bias than that of the contacts, and the polarity reversal is reproduced in the simulation (Fig. 3c,d). From this comparison, we deduce that the applied bias has a stronger influence on the band bending at the 1L/ML junction than at the contacts, as confirmed directly by the calculated band diagrams of Figure 4a. The simulations also generate the rectifying *I-V* characteristic that is observed in experiments (Supp. Fig. S4) o*nly* when the 1L electron affinity is *greater* than that of the 3L. With this type-II band alignment (Fig. 3d), the potential drop at the 1L/ML junction is larger under positive drain voltages than under negative drain voltages according to the simulations. Experimentally, this difference was confirmed in electrostatic force microscopy (EFM) potential profiles along a 1L/ML device (Supp. Fig. S5). The degree of agreement is perhaps surprising given that the simulations, by assuming a 3-D density of states with a uniform in-plane mobility and effective mass, do not account for quantum effects that one might assume to be important in monolayer thick materials. However, we can conclude that the simulations capture the predominant behavior of free carriers generated by photoexcitation in these devices, with the caveat that additional effects might be observed at low temperatures in suspended films of extremely high quality, for which phonon, interface, and defect scattering are minimized. One can further infer that the behavior of free carriers in these devices is most sensitive to variations in carrier concentration and electron affinity differences at the junction.

A parameter sensitivity analysis was conducted for both carrier concentration and electron affinity to probe the sensitivity of SPCM measurements to the nature of the heterojunction. For reasonable ranges of the free carrier concentrations, the SPCM response could only be reproduced by a narrow range of electron affinities with $\chi_{1L} > \chi_3$ (Fig. 4), which corresponds to a type II band alignment. Specifically, $\chi_{1L}$ must be large enough, and the difference between $\chi_{1L}$ and $\chi_{3L}$ sufficient, to generate the observed relative photocurrent magnitudes at the contact junction and the discontinuous thickness junction. The results are insensitive to the magnitude of the 1L gap in simulations for values as low as 2.1 eV (which lies closer to the "optical gap"), reflecting the apparent dominance of the conduction band alignment in controlling the photoresponse (the valence band offset is always large). A review of first principles calculations of $MoS_2$ electron affinities does not reveal a consensus on whether the band alignments of isolated (i.e., substrate-free) 1L-ML junctions are type-I or type-II. Some calculations have found a $\chi_{1L}$ ~4.27 eV, [30, 39] which is larger than bulk (~4 eV)[32] and thus consistent with our findings. However, other studies have proposed a small type-I offset[29, 40] to explain the dependence of contact resistance on layer number.[40] The limitations of first-principles calculations of band edges are well known, and indeed it appears that prior work neither confirms nor refutes our interpretation of the band alignment.

Limits of the experimental approach should also be considered. For example, the deduced affinity offset between the monolayer and multilayer regions could be sensitive to extrinsic factors including fixed charge at the $SiO_2$-$MoS_2$ interface and surface adsorbates, since the experiments were conducted under ambient conditions. In particular, adsorbates could form surface dipoles that modulate the electron affinities of monolayer and multilayer regions to different degrees[41] because any trapped charge at the interface would be screened to a lesser degree in the monolayer region. Adsorption at the step edge could also modify the band profile in the heterojunction. For this reason, it is desirable to extend these investigations to vacuum and/or encapsulated junctions in the future.

As shown above, the qualitative and quantitative correspondence between experiment and simulations support our interpretation that the photocurrents at 1L/ML junctions and ohmic contacts are dominated by charge carrier separation induced by electric fields when the device is under bias. Interestingly, the comparatively smaller junction photoresponse at zero bias is positive in experiments (Fig. 3b, 5a), whereas the simulation shows that the negative photovoltaic response should dominate the positive photothermoelectric contribution to the photocurrent (Fig. 3c, 5b). Furthermore, the zero-bias response is only weakly dependent on gate voltage (Supp. Fig. S6). Apparently, either the photothermoelectric contribution is greater (even more positive) than expected, or an additional generation/transport mechanism occurs at the junction that is not captured by the modelling of thermally equilibrated, Fermi-Dirac distributed charge carriers. Consistent with this second hypothesis, the spectral response at the discontinuous thickness junction is distinct from the energy dependence of the photovoltaic contact response (Fig. 5c), which follows the optical absorption profile of $MoS_2$. Below we explore non-exclusive possibilities: (1) the extent to which an extraordinary Seebeck effect could explain the response, and (2) the extent to which "hot" carriers, which have not equilibrated with the lattice temperature, may contribute to the photocurrent.

To clearly define and discriminate between possible contributions, we refer to photoinduced currents driven by a carrier temperature gradient as photothermal in nature. Photothermal currents include both photoinduced thermoelectric currents (in which charge carrier and lattice temperatures are equal) and hot carrier transport[42, 43] (in which the charge carrier temperature is higher than the lattice temperature) such as in internal photoemission at junctions[44]. Photothermoelectric currents, for example, have been reported at 1L and 2L graphene junctions[25] and at monolayer steps in topological insulators[26], geometries that are similar to that considered here. A photothermoelectric response is also present in our MoS$_2$ 1L/ML junctions because the valley degeneracy in the conduction band of the multilayer (6) is three times that of the monolayer (2), leading to a discontinuity in the density of states and therefore the Seebeck coefficient at the 1L/ML junction. Indeed, the experimental photocurrents observed in all devices studied are consistent with a flow of electrons into the thicker MoS$_2$ material upon photoillumination. However, the Seebeck coefficients used in our simulations are too small to enable the *positive* photothermoelectric contribution to dominate the *negative* photovoltaic contribution. Simpler approximate analytic solutions reach the same conclusion: we approximate the Seebeck effect at the junction as $\Delta S/\Delta T = I_{ph}*R$,[26] where the difference in temperature between the junction and the contacts $\Delta T < 9$ K is calculated with a finite element model, and the resistance $R > 1$ G$\Omega$ is determined from *I-V* curves. We find the difference in Seebeck coefficient between the 1L and ML materials $\Delta S$ must be greater than 900,000 µV/K to produce the observed I$_{ph} \sim 0.1$ nA. In contrast, a recent measurement of monolayer CVD MoS$_2$ in vacuum[45] finds $\Delta S$ to be at least 1-2 orders of magnitude lower than this value, and analytical estimates using S$= -\frac{k_B}{q}\left[3 + \ln\frac{N_c}{n}\right]$, where $N_c$ is the effective density of states in the conduction band and $n$ is the carrier concentration[46], are approximately 4 orders of magnitude smaller, as are theoretical predictions.[47] Finally, a *conventional* photothermoelectric effect is definitively ruled out by the spectral photoresponse of the junction (Fig. 5c), which does not reflect the absorption versus energy of either SL or 3L MoS$_2$. Rather, the onset of the junction photocurrent (Fig. 5c) occurs at a higher energy than photocurrent at the contacts, suggesting that electrons with energy beyond that of the gap (i.e. hot electrons) are responsible for the positive photothermal current.

We consider two sources of hot carrier photocurrent. First, we note that we have previously demonstrated the capability to measure the internal photoemission of hot electrons at a nanoscale metal-semiconductor junction[48]. In this prior work, hot electrons travel ballistically from the photoexcited metal into the semiconductor collector with an energy dependent yield given by the Fowler model. In the present case, assuming that free carriers rather than excitons dominate the ultrafast response at our fluence (~0.5 mJ/cm$^2$) as shown in reference[49], we estimate that $\sim 1.5 \times 10^5$ electrons are generated per pulse. Using an electron Fermi velocity of $3 \times 10^7$ cm/s[50, 51] and a thermalization rate of 14 fs, we estimate that ~170 electrons per pulse within the momentum escape cone reach the junction without scattering. The observed current of 0.1 nA would then require an injection efficiency of only 5.5% (9 electrons/pulse), corresponding to an internal (hot carrier) quantum efficiency of ~0.003%. This is well within previously observed limits of internal quantum efficiencies for hot carrier injection of 0.1-5% for internal photoemission at metal-semiconductor junctions[52, 53]. Second, the band bending near the junction could assist in injection provided the interaction time is sufficiently long for carriers to acquire momentum from the electric field. Therefore, thermally assisted tunneling of drifting carriers could also contribute to the hot electron photocurrent during the time period required for the equilibration with the lattice (0.6 ps)[49]. As the electron-phonon equilibration time is approximately 40 times longer than the electron-electron thermalization time, substantially more carriers could contribute to the current than for direct ballistic injection. The quantitative analysis of experimental carrier injection above provides a basis for more sophisticated time-dependent calculations that are beyond the scope of the present work.

In summary, we report the formation of rectifying heterojunctions at the interface of monolayer and multilayer MoS$_2$. Experiments and finite element modeling of local photocurrent generation suggest a type II band alignment, and spectroscopic analysis indicates that hot electrons contribute to the photoresponse. The quantitative analysis of photoresponse mechanisms and their sensitivity to band alignment contribute to the understanding of heterostructures in the ultimate size limit, thereby supporting the rational design of future optoelectronic 2-D devices. With the availability of CVD grown MoS$_2$ and fabrication techniques, including laser thinning[54], layer-by-layer etching[55], and layer stacking[34], the 1L/ML junction concept can be readily explored and potentially scaled to larger areas.

**Acknowledgements:**

This work was primarily supported by the Materials Research Science and Engineering Center (MRSEC) of Northwestern University under National Science Foundation (NSF) Grant DMR-1121262. S.L.H. acknowledges support in part by the NSF Graduate Research Fellowship under DGE-1324585. L.J.L. acknowledges support from DOE BES grant DE-FG02-07ER46401 for S.L.H., SPCM instrumentation, and analysis software. D.J. acknowledges support in part by the SPIE Optics and Photonics Graduate Scholarship. This research made use of the Northwestern University's Atomic and Nanoscale Characterization Experimental Center at Northwestern University, which is supported by the NSF Nanoscale Science and Engineering Center, NSF-MRSEC (DMR-1121262), Keck Foundation, and State of Illinois.

**Author contributions statement:**

D.J. and S.L.H. contributed equally to this work. C.C.W. and S.L.H. performed SPCM and Raman measurements. D.J. fabricated devices and performed electrical characterization. S.L.H. did the modeling and co-wrote the manuscript with L.J.L. K.S.C. did the EFM measurements. V.K.S. and J.K. assisted with device fabrication. C.C.W., D.J., and L.J.L. collaborated on the experimental design. All authors discussed the work and participated in completion of the manuscript.

**Supporting Information**: Includes modeling details and supplementary simulation and measurement figures. This material is available free of charge via the Internet at http://pubs.acs.org.

The authors declare no competing financial interest.

**Figures**

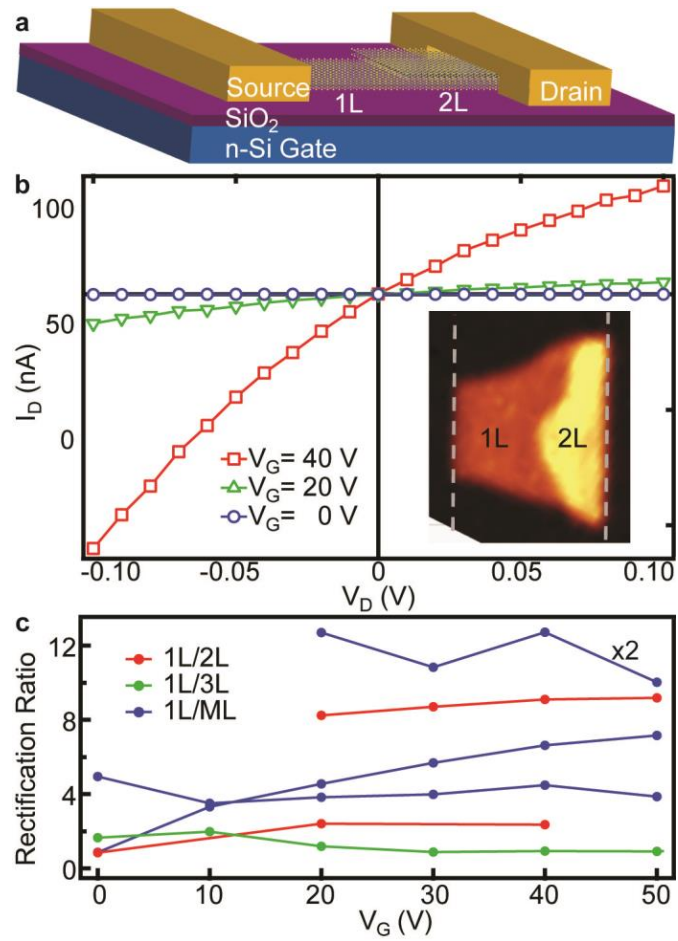

**Figure 1:** (a) Schematic side view of a 1L/2L MoS$_2$ junction FET. (b) Drain-source output curves under gate biases V$_G$= 0 V, 20 V, and 40 V. Inset of (b): Raman map of integrated intensity between 382 and 385 cm$^{-1}$. The dotted lines indicate source and drain contacts separated by 4.4 μm. (c) Rectification ratio of 1L/ML devices as a function of gate bias.

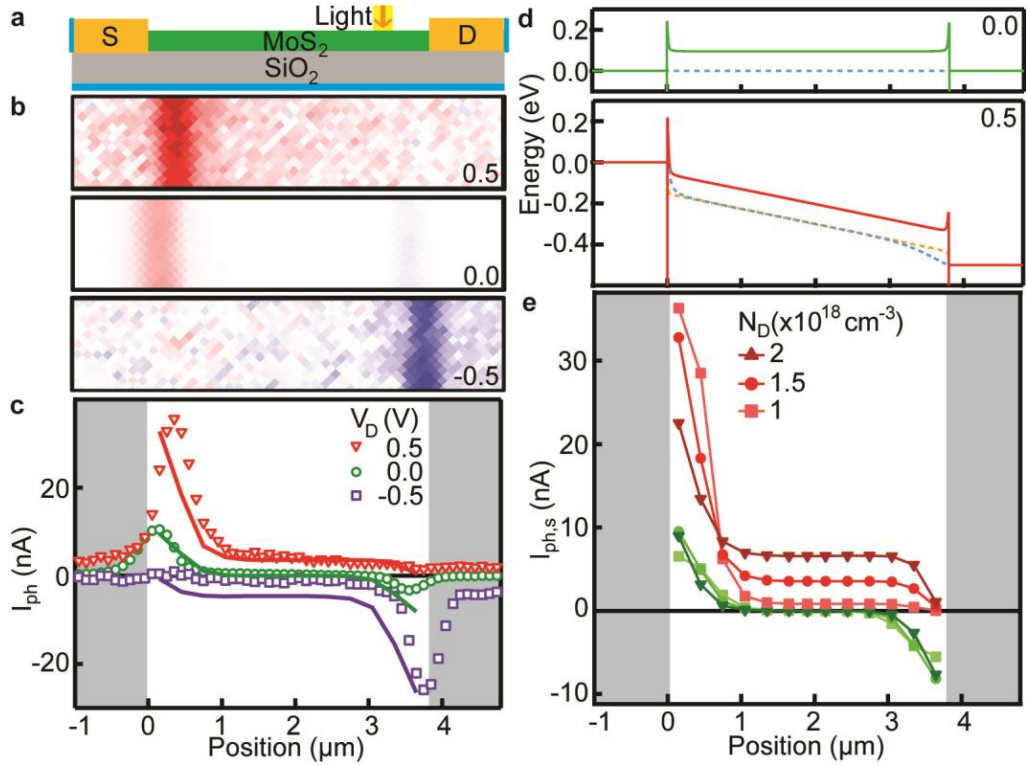

**Figure 2:** (a) Schematic of SPCM measurement on a uniform thickness 4L device. (b) SPCM image at 0.5, 0, and -0.5 $V_D$ (top to bottom) and 550 nm. (c) Measured (open shapes) and simulated (solid lines) photocurrent profiles under bias. Gray areas indicate contact locations. Experimental data in (b,c) are reproduced from Wu, et al.[1] (d) Simulated band diagrams at 0 and 0.5 $V_D$ (top, bottom) depicting the conduction bands and Fermi levels. (e) Sensitivity of simulated SPCM profiles to carrier concentration. Green (red) curves at 0.0 (0.5) $V_D$.

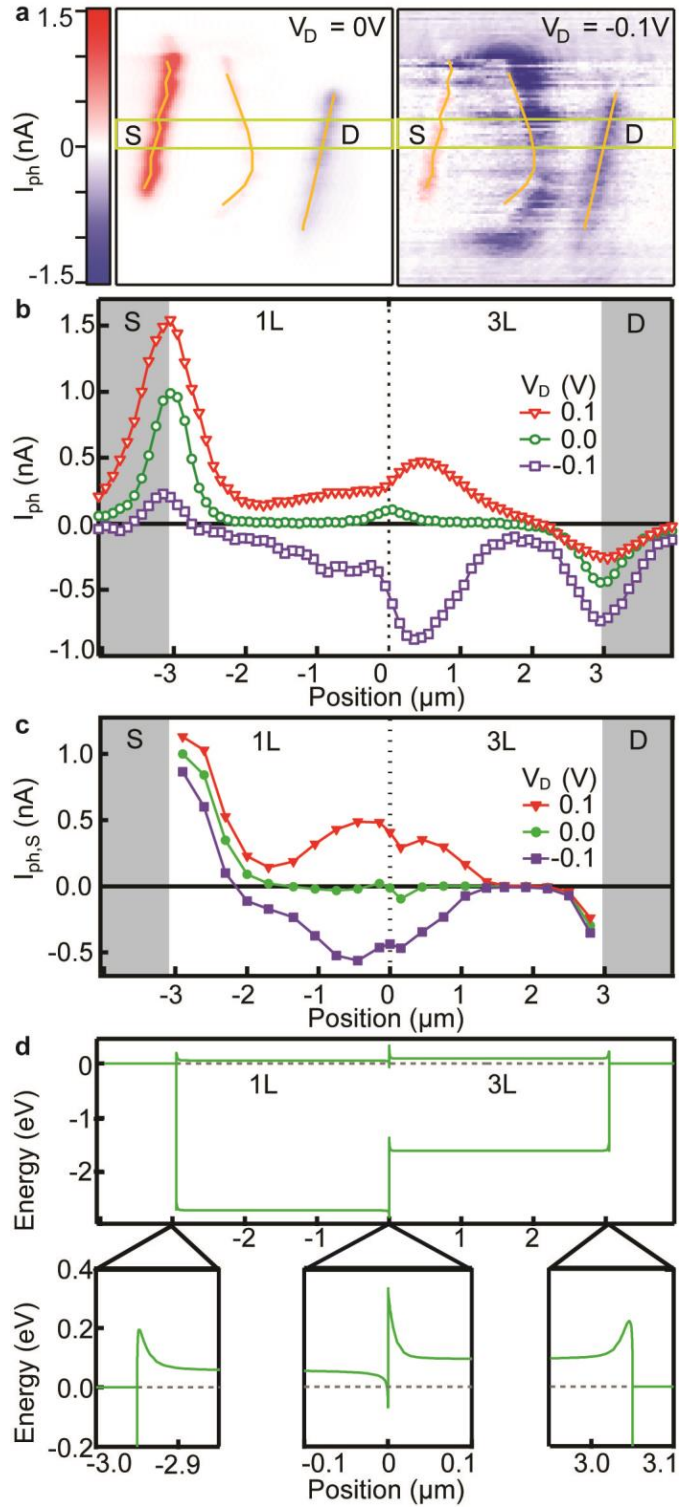

**Figure 3:** (a) Spatial photocurrent mapping of a 1L/3L device at $V_D$= 0V and -0.1V ($V_g$=0, $\lambda$=525 nm). Yellow lines indicate approximate positions of the contacts and 1L/3L junction. (b) Photocurrent profiles from the region of the green boxes in (a). Profiles at $V_D$ = 0.1, 0.0, and -0.1 V. The gray areas and dotted line indicate contact locations and 1L/3L junction, respectively. (c) Simulated photocurrent profiles for the experimental conditions in (b). (d) Simulated band diagram at equilibrium ($V_D$ =0 V).

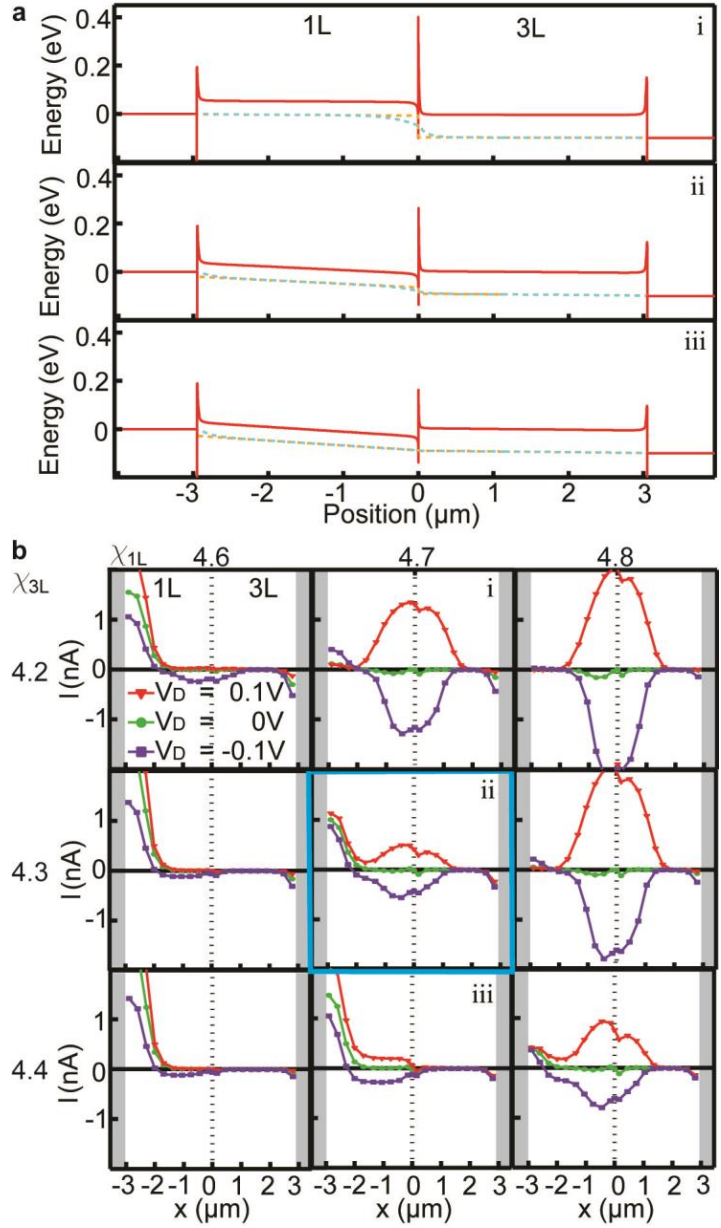

**Figure 4:** (a) Simulated 1L/3L conduction bands and quasi-Fermi levels when $V_D = 0.1$ V with constant electron affinity $\chi_{1L}$=4.7 eV and varying $\chi_{3L}$ = 4.2, 4.3, and 4.4 eV in i, ii, and iii respectively. (b) Simulated photocurrent versus position $x$ at varying electron affinities. The gray areas and dotted line indicate contact regions and 1L/3L junction, respectively.

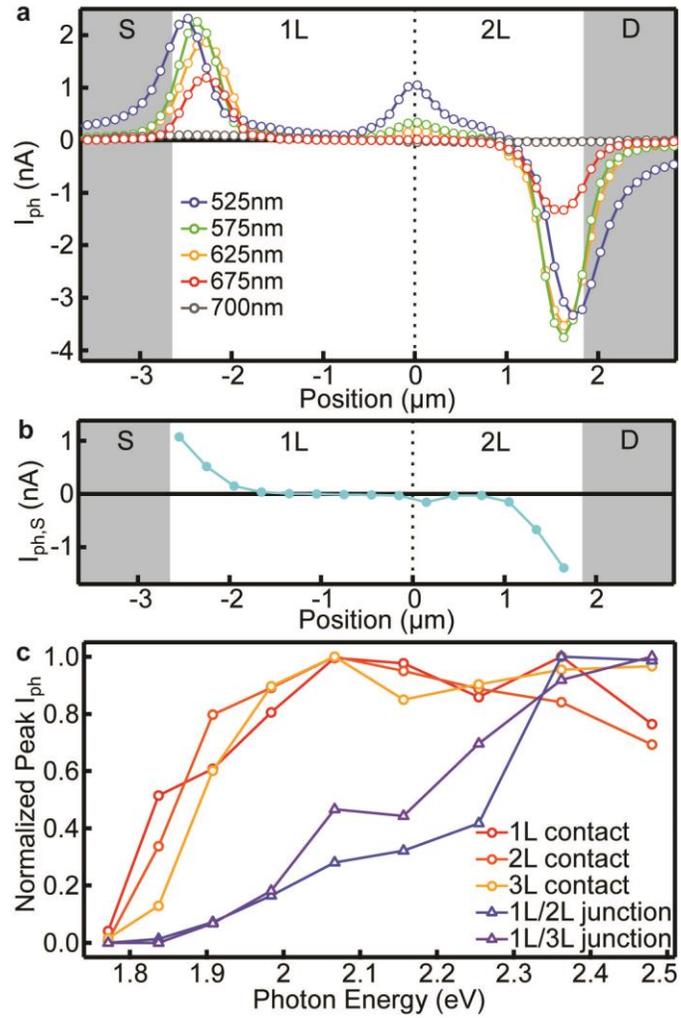

**Figure 5:** (a) Zero-bias experimental SPCM profiles of a 1L/2L device at different excitation wavelengths. Gray areas indicate contact regions. Positive current flows from drain to source. (b) Simulated SPCM profile at 500 nm excitation. (c) Spectral dependence of short-circuit photocurrent for local illumination at the contacts (red, orange, yellow) and at single layer-multilayer junctions (blue, purple).

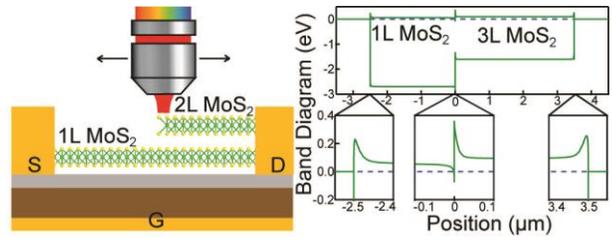

**Figure TOC graphic**

# Supporting Information:

## Investigation of Band-Offsets at Monolayer-Multilayer MoS₂ Junctions by Scanning Photocurrent Microscopy

Sarah L. Howell[§], Deep Jariwala[§], Chung-Chiang Wu, Kan-Sheng Chen, Vinod K. Sangwan, Junmo Kang, Tobin J. Marks, Mark C. Hersam, and Lincoln J. Lauhon

### Simulation Details

Sentaurus TCAD was used to model MoS₂ devices by solving the following steady-state coupled differential equations in 2 dimensions:

$\nabla \cdot (\varepsilon \nabla \varphi) = -q(p - n + N_D - N_A)$   Poisson's equation

$\nabla \cdot J_n = q^* R_{net}$   Electron continuity equation

$-\nabla \cdot J_p = q^* R_{net}$   Hole continuity equation

$J_n = -n \cdot q \cdot \mu_n (\nabla \Phi_n + P_n \nabla T)$   Electron current equation

$J_p = -p \cdot q \cdot \mu_p (\nabla \Phi_p + P_p \nabla T)$   Hole current equation

$-\nabla \cdot \kappa \nabla T = M$   Temperature equation

where $\varepsilon$ is the electrical permittivity, $\varphi$ is the electrostatic potential, $q$ is the charge of an electron, $p$ and $n$ are the hole and electron densities, $N_D$ and $N_A$ are the donor and acceptor concentrations, $J_n$ and $J_p$ are the electron and hole current densities, $R_{net}$ is the net recombination rate (including the generation rate contribution), $\mu_n$ and $\mu_p$ are the electron and hole mobilities, $\Phi_n$ and $\Phi_p$ are the electron and hole quasi-Fermi potentials, $P_n$ and $P_p$ are the absolute thermoelectric powers, $\kappa$ is the thermal conductivity, $T$ is the lattice temperature, and $M$ is a function for the total heat generation rate.

The geometry is shown in Figure 2a inset. When available, the materials parameters used in the simulations were taken from the literature. Field effect mobilities and free carrier lifetimes were constrained by experimental measurements on uniform thickness FETs. Simulations of 1L/ML junctions used parameters that were optimized by fitting the electrical characteristics and SPCM profiles of uniform thickness FETs. Three-dimensional effective densities of states were calculated using conduction (valence) band valley degeneracies of 2 (2) and 6 (2) for 1L and multilayers, respectively.[8, 9] Three-dimensional effective density of state equations are used, Fermi-Dirac carrier statistics were used, and carrier concentrations were specified by specifying a doping level and assuming complete ionization. The mobility is assumed to be independent of carrier density. Since the anisotropic (in-plane vs. cross-plane) mobility in MoS₂ has not been experimentally measured, for simplicity we assume isotropic transport mobility. A Shockley-Read-Hall (SRH) recombination model was implemented with free-carrier lifetimes constrained by experiments.[12] Generation due to photoillumination was simulated in 300 nm rectangular windows similar in size to the FWHM of the focused laser beam. We assume a quantum yield of 1% to account for exciton dissociation. For simplicity, we use a spatially uniform quantum yield, while noting that additional excitons dissociate into free carriers when the electric field is large near junctions[14]. We assume transport through the contacts is ohmic (charge neutrality and equilibrium are assumed) and assume no Fermi level pinning at the metal contacts. Ideal heat sinks are imposed at the outside edge of each contact with the Dirichlet boundary condition, T = 300K. Transport though the single layer-multilayer interface junctions is governed by a thermionic emission model with WKB tunneling.

**Table S1:** The material constants that were used in the simulations unless otherwise noted.

| Parameter | Symbol | Value | References and Comments |
|---|---|---|---|
| Thickness of MoS₂ | | 0.7nm/layer | |
| Band gap 1L | $E_g$ | 2.76 eV | [9, 15, 16] Note: The transport gaps are larger than the optical band gaps due to large exciton binding energies. |
| Band gap 2L | | 1.9 eV | |
| Band gap 3L | | 1.7 eV | |
| Band gap 4L | | 1.6 eV | |
| Electron affinity 1L | $\chi$ | 4.7 eV | Bulk value is ~4eV.[17] The trend was determined by parameter sweep fits (Figure S3, S5). |
| Electron affinity 2L | | 4.4 eV | |
| Electron affinity 3L | | 4.3 eV | |
| Electron affinity 4L | | 4.0 eV | |
| Dopant concentration | $N_D$ | 1.5e18/cm³ | Sheet concentrations are generally found between $10^{10}$ and $10^{12}$/cm², corresponding to $4 \times 10^{16}$ to $1.4 \times 10^{19}$/cm³. |
| Permittivity 1L | $\varepsilon$ | 4.2 $\varepsilon_o$ | [16] |
| Permittivity 2L | | 6.4$\varepsilon_o$ | |
| Permittivity 3L | | 8.8 $\varepsilon_o$ | |
| Permittivity 4L | | 11 $\varepsilon_o$ | |
| Electron DOS mass SL | $m_e$ | 0.5 $m_o$ | [15, 16] |
| Electron DOS mass FL | $m_e$ | 0.5 $m_o$ | |
| Hole DOS mass SL | $m_h$ | 0.5 $m_o$ | |
| Hole DOS mass FL | $m_h$ | 1 $m_o$ | |
| Mobility 1L | $\mu_L$ | 10 cm²/V/s | The field effect mobility of 1L, 2L, 3L, and 4L FET devices ranges from 1 to 10, 4 to 20, 5 to 30, and 8 to 30 respectively, when measured under ambient conditions. |
| Mobility 2L | | 20 cm²/V/s | |
| Mobility 3L | | 30 cm²/V/s | |
| Mobility 4L | | 30 cm²/V/s | |
| SRH lifetimes 1L | $\tau$ | 0.5-1.5 ns | Free carrier lifetime in ML materials is on order 0.1ns[12] and we expect 1L lifetime to be longer than this with its direct band gap. We use a lifetime slightly higher than this particular experimental value so that the simulated photocurrent decay away from contacts (related to the mobility lifetime product) agrees better with our SPCM measurements. |
| SRH lifetimes 2L | | 0.5 ns | |
| SRH lifetimes 3L | | 0.3 ns | |
| SRH lifetimes 4L | | 0.3 ns | |
| Quantum Yield 1L | QY | 1-2% | Since these materials have a large exciton binding energy, we expect the QY (the fraction of absorbed photons that create free electron hole pairs) to be less than 100%. |
| Quantum Yield 2L | | 1% | |
| Quantum Yield 3L | | 1% | |
| Quantum Yield 4L | | 3% | |
| Workfunction of gate | | 4.05 eV | Electron affinity of silicon. |
| Workfunction of gold | | 5.1 eV | |
| 1L Absorption at 500nm | | 10% | [18] Each layer absorbs ~10% of the incident light at 500nm. |
| 2L Absorption at 500nm | | 25% | |
| 4L Absorption at 500nm | | 47% | |
| 1L Absorption at 525nm | | 8.5% | |
| 3L Absorption at 525nm | | 30% | |
| Lattice Heat Capacity | $c_L$ | 2 J/K²/cm³ | [19, 20] Value for single layer. 400 J/kg/K and 5000 kg/m³ |
| Thermal conductivity | $\kappa$ | 50 W/mK | [21, 22, 23] Simulations are insensitive to this value, since most of the heat is lost through the substrate. |

Notes: All constants used for SiO₂ were the Sentaurus library values.[24-26]

Seebeck coefficients are calculated using a simple analytical model[24-26]:

$$P_n = -\kappa_n \frac{k}{q}\left[\left(\frac{5}{2} - s_n\right) + \ln\left(\frac{N_C}{n}\right)\right] \text{ and } P_p = \kappa_p \frac{k}{q}\left[\left(\frac{5}{2} - s_p\right) + \ln\left(\frac{N_V}{p}\right)\right],$$ where $S_n$ and $S_p$ are -0.5 and $\kappa_n$ and $\kappa_p$ are 1.

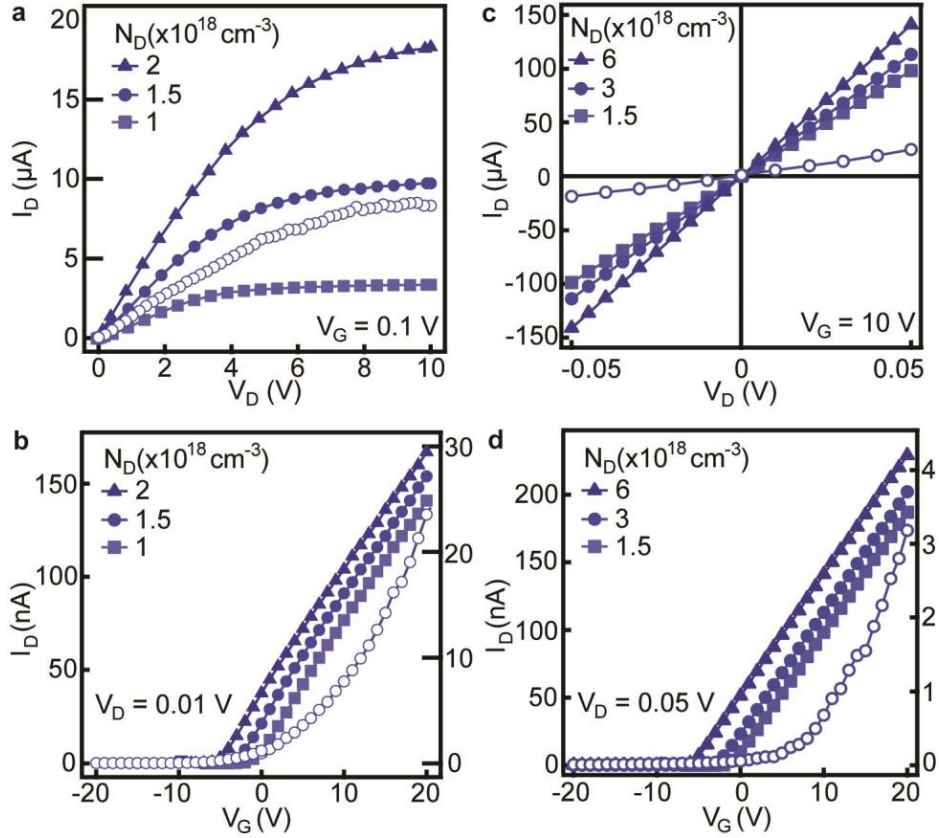

**Figure S1:** Simulated (solid markers, left axis) and experimental (empty markers) (a) output characteristics when $V_G$ = 0.1 V and (b) transfer characteristics when $V_D$ = 0.01 V for a 4L thick FET. Simulated (solid markers, left axis) and experimental (empty markers) (c) output characteristics when $V_G$ = 10 V and (d) transfer characteristics when $V_D$ = 0.05 V for a 1L thick FET. Experimental data in are reproduced from Wu, et al.[1] The doping concentration of 1.5e18cm$^{-3}$ yields output, transfer, and SPCM profiles (Figure S2) that agree well with experiments. Ideal transistor transfer curves show zero current below the threshold voltage and a linear gate bias dependence (i.e. a constant slope) above the threshold voltage. Deviations of experimental transfer curves from ideal simulations have been attributed to scattering from unscreened charged impurities[5] and a carrier density-dependent mobility[10]. To elaborate, mobility limited by Coulomb scattering from unscreened charged impurities and/or defects at the $MoS_2$-$SiO_2$ interface results in a supralinear dependence of conductivity on carrier concentration (within this bias range at room temperature). This effect leads to approximately parabolic instead of linear experimental transfer curves above the threshold voltage and reduces current magnitudes. An additional series contact resistance may further limit the current magnitudes in the experimental devices.

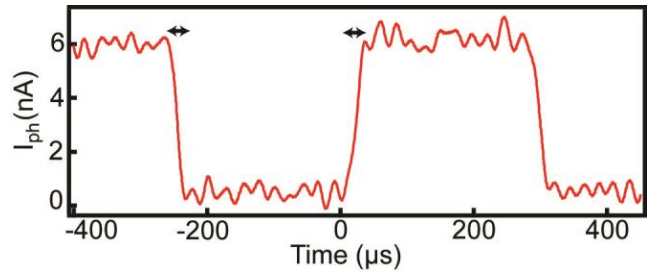

**Figure S2:** Time-resolved photocurrent of a 1L-ML MoS₂ FET with an illumination on/off modulation frequency of 1837 Hz recorded with current preamplifier (DL Instruments 1211) and a digital sampling oscilloscope (Tektronix TDS 2014). The rise and fall times (10% to 90%) of the photocurrent are faster than 30 µs. As discussed previously in the supplement of Wu, et al.[1], the photocurrent response is limited by the current preamplifier bandwidth. The time-dependence of the photocurrent is consistent with a fast process such as interband excitation and charge carrier separation.

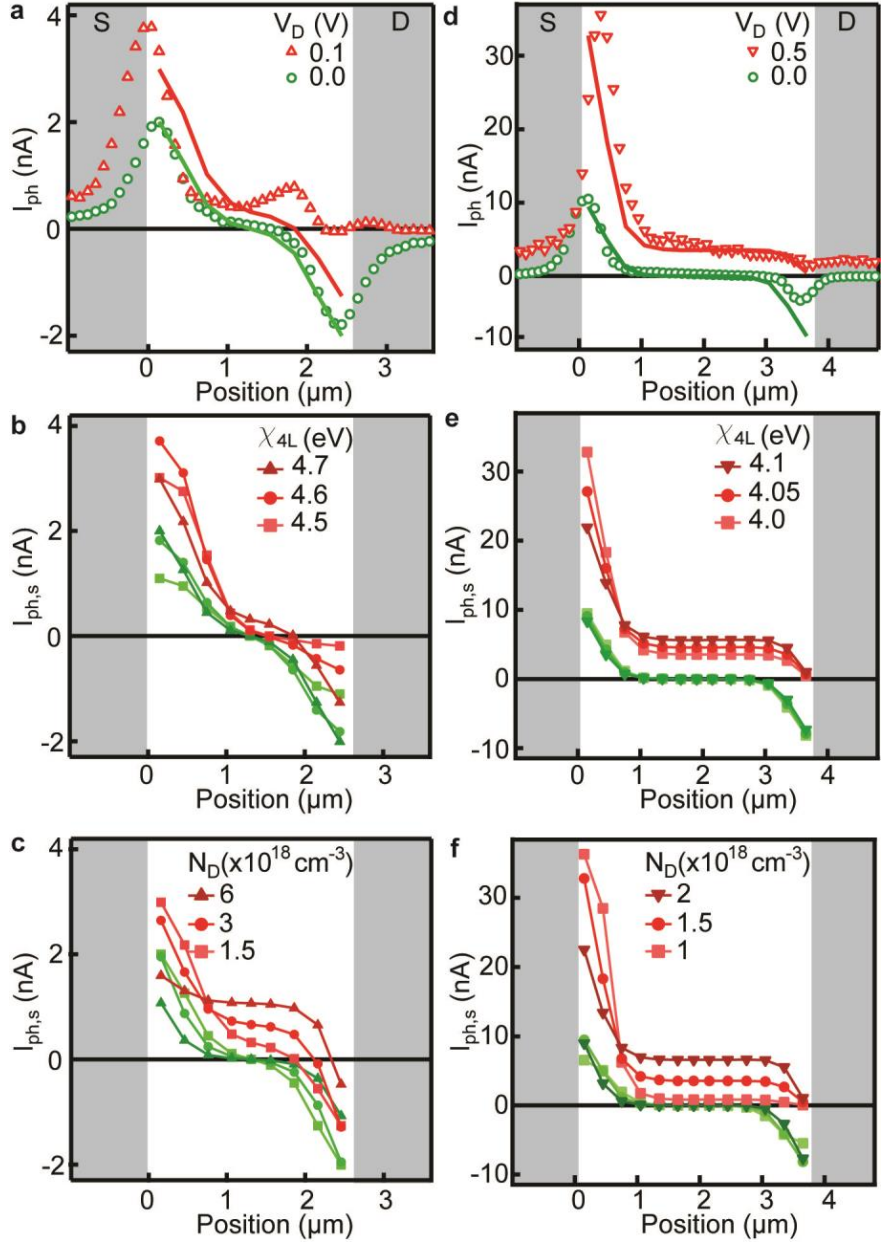

**Figure S3:** (a) Measured (open shapes) and simulated (solid lines) photocurrent profiles under bias for a 1L FET. Gray areas indicate contact locations, with the drain contact at the right. Sensitivity of simulated SPCM profiles to (b) electron affinity and (c) carrier concentration for a 1L FET. Green (red) curves at zero (positive) $V_D$. (d-f) are the same as (a-c) for a 4L FET. Experimental data in (a) and (d) are reproduced from Wu, et al.[1, 2] The doping concentration of $1.5e18cm^{-3}$ and electron affinities of 4.0 eV (4.7 eV) for 4L (1L) FETs yields output, transfer (Figure S1), and SPCM profiles that agree well with experiments.

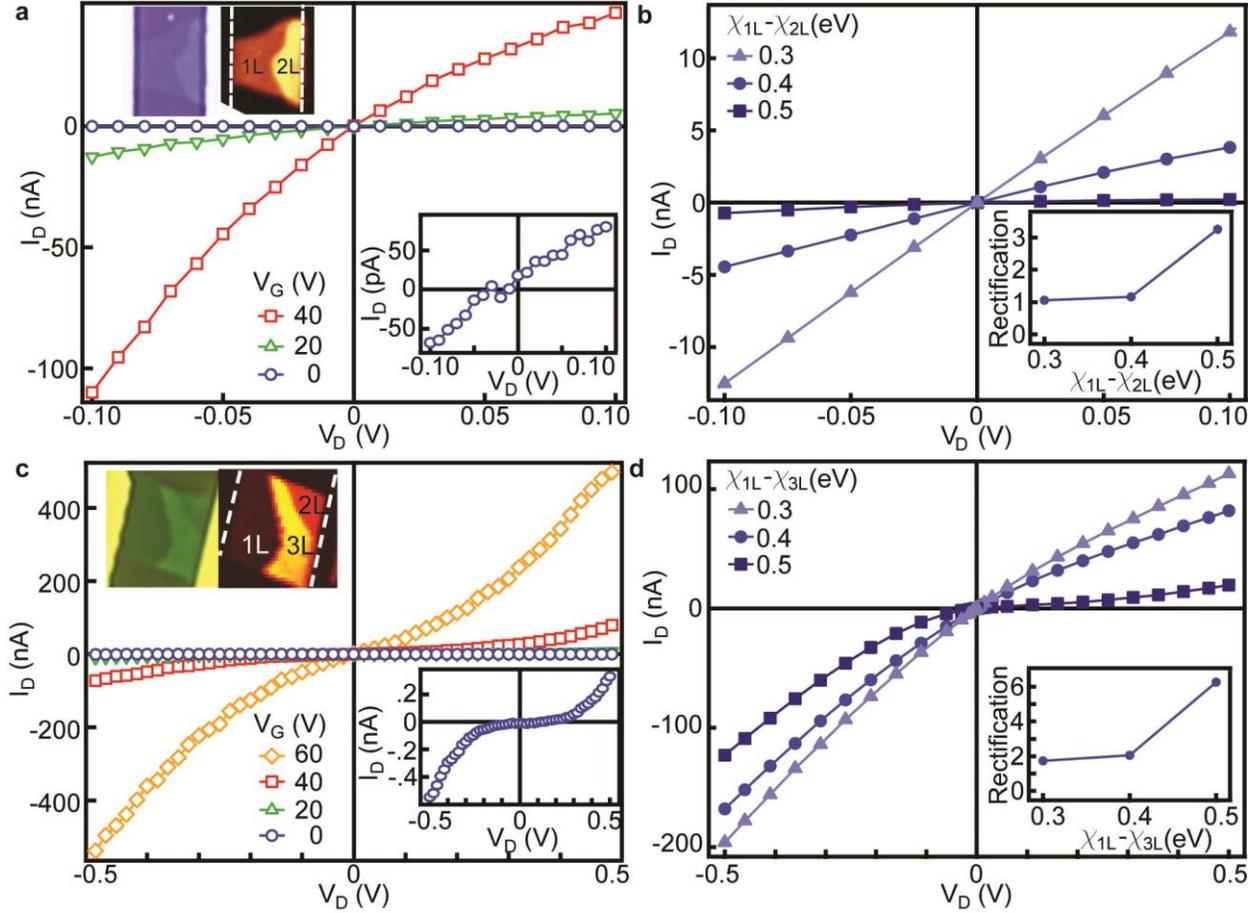

**Figure S4:** (a),(c) Experimental drain-source output curves of (a) a 1L/2L device and (c) a 1L/3L device under gate biases $V_G$= 0V, 20V, 40V, and 60V. The top left insets of (a) and (c) are optical microscopy images and Raman spectroscopy maps with the MoS$_2$ thicknesses labeled. The dotted lines indicate source and drain contacts separated by 4.4 μm and 6 μm respectively. The bottom right insets of (a), (c) replot the measured output curves at $V_G$= 0V. (b),(d) Simulated output curves at $V_G$= 0V of (b) 1L/2L device and (d) 1L/3L device. The insets of (b),(d) plot the rectification ratio as a function of the difference in electron affinity. (b),(d) show that the nonlinear, weakly rectifying behavior is reproduced in the simulations and originates from the 1L/ML junction, increasing in rectification as the difference in electron affinity is increased (here $\chi_{1L}$ = 4.7). While the trend in rectification is reproduced in the simulations, the simulated current magnitudes are larger than those measured. The reduced experimental current magnitudes are attributed in part to series resistance at the contacts, which is sensitive to factors not included in the model, such as impurities[3], specific contact geometries[6], the orbital overlap at the junction[11], and Fermi-level pinning[13].

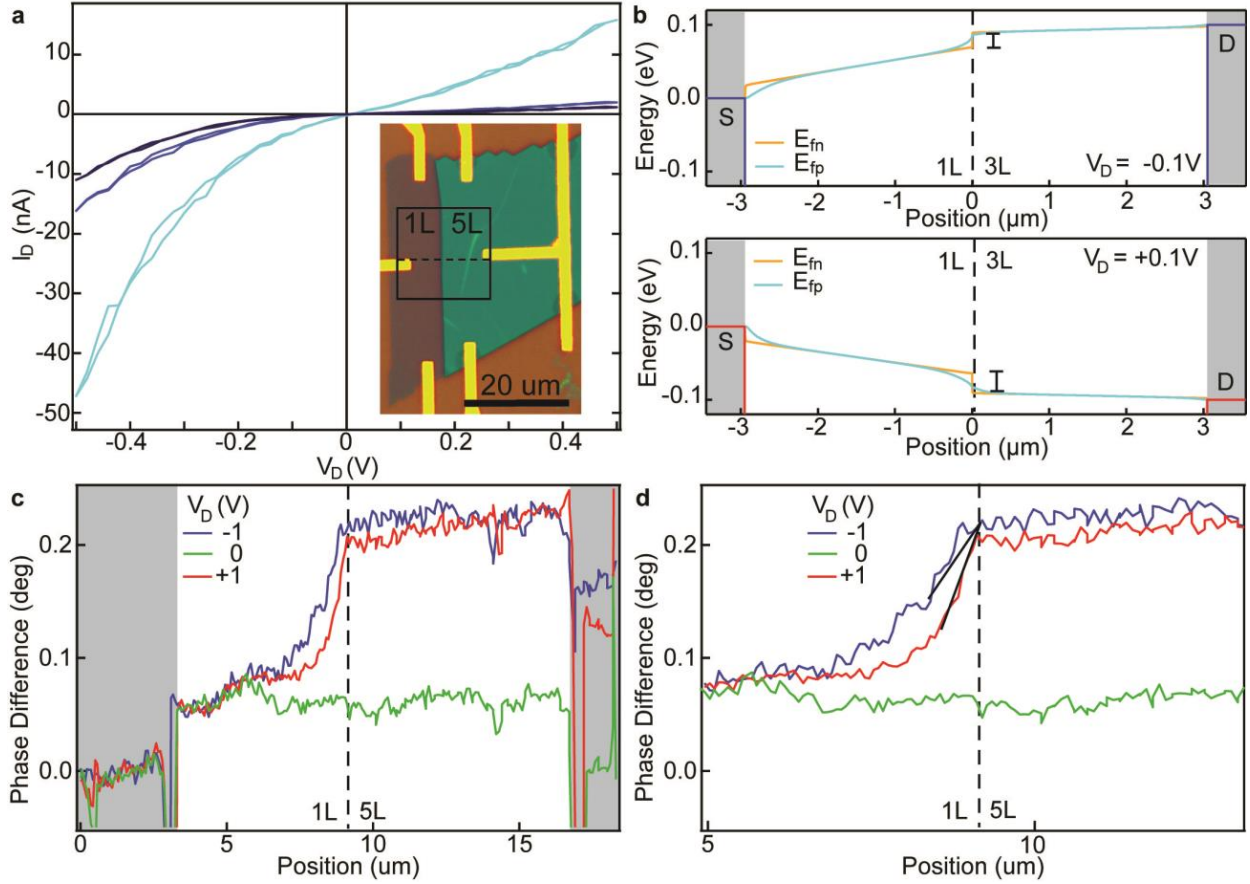

**Figure S5:** (a) *I-V* curves at 0 V$_G$ for three pairs of electrodes across the 1L(source)/5L(drain) device shown in the inset optical image. (b) Simulated potential profiles for a 1L/3L device as described in the manuscript. (c) Electrostatic force microscopy (EFM) phase shift profiles for +1, 0, and -1 V$_D$ taken along the dotted line in the inset of (a). (d) EFM phase shift profiles for +1, 0, and -1 V$_D$ for the region near the 1L/5L junction; black lines are guides to the eye.

The *I-V* curves shown in Fig. S5a for a 1L/5L device exhibit rectification in the same direction as the examples in the main text, namely, the device is more (less) conductive when the multilayer drain contact is negatively (positively) biased. Simulations of a 1L/3L device (Fig. S5b, Fig. 3) under forward (top) and reverse (bottom) bias show that this difference in conduction arises primarily from an increase in the space charge region in the 1L material in the vicinity of the 1L/3L junction under reverse bias. Hence, a larger fraction of the potential drop occurs within this region under reverse bias compared to forward bias.

Electrostatic force microscopy (EFM) profiles under forward and reverse bias (Figs S5c,d) of the device shown in Fig. S5a confirm this change in band profiles. In EFM, a conductive atomic force microscope cantilever is raster scanned at a fixed height (tens of nm) above the sample, while the phase shift of the oscillating cantilever is simultaneously recorded. The measured phase shift $\Delta\varphi$ shown in (c) is proportional to the square of the difference in surface potential between tip and sample:

$$\Delta\varphi \propto \frac{Q}{2k}\frac{\partial^2 C}{\partial z^2}[V_T - V_S + \Delta\phi]^2$$

where $Q$ and $k$ are the quality factor and the spring constant of the cantilever respectively, $\frac{\partial^2 C}{\partial z^2}$ is the second partial derivative of the tip-sample capacitance with respect to the tip-sample distance $z$, $V_S$ and $V_T$ are the sample and tip bias, respectively, and $\Delta\varphi$ is the work function difference between the tip and sample.[27, 28] Here $V_T$ was always grounded. The phase increases upon scanning from the metal contacts to the MoS$_2$ channel due to the decrease in work function and corresponding increase in surface potential difference (Fig. S5c, green curve). Fluctuations due to noise and topographical artifacts dominate the surface potential difference between the 1L and 5L regions at zero bias. Under non-zero drain bias, there is virtually no change in the potential across the 1L-MoS$_2$ contact, as shown by the overlapping green, red, and blue curves (Fig. S5c, left). However, there is a steep phase (and therefore potential) drop in the 1L region immediately adjacent to the junction, with a more gradual potential decrease through the remainder of the 1L region, consistent with the simulations. The difference between the device response at forward and reverse bias is more clearly seen in Fig. S5d, for which black lines have been drawn as guides to the eye. The reverse (positive) bias induces a steeper potential drop in the 1L region adjacent the junction, again consistent with simulations. Furthermore, there is negligible change in the phase of the MoS$_2$ immediately adjacent to the contacts. Thus, EFM shows the junction to be more (less) conductive when the ML drain contact is negatively (positively) biased, which would result in the direction of rectification that we observe. Hence, we conclude that the rectification arises from the 1L/5L junction.

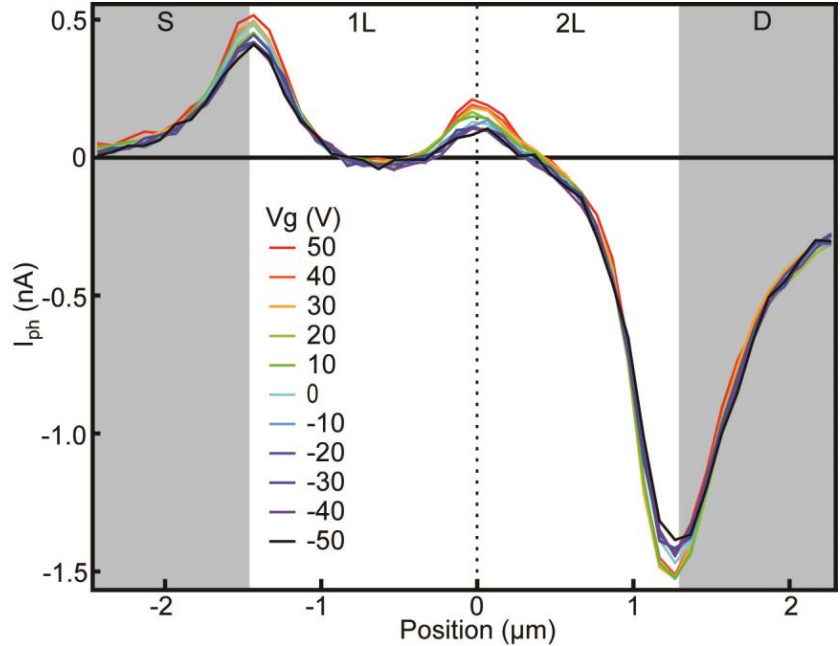

**Figure S6:** (a) Photocurrent profiles at 0 $V_D$ for a 1L/2L device showing dependence on gate bias. The dotted lines indicate source and drain contacts. The photocurrent under local illumination has a modest gate dependence within this regime of gate biases. The small increase in photocurrent magnitude at the metal/1L and 1L/2L junctions with increasing gate bias are discussed separately below.

As discussed in the main text, the photocurrent at the 1L/2L junction is dominated by hot carriers that are not in equilibrium with the lattice. For example, one can expect a contribution from internal photoemission, in which hot electrons in the 1L material are injected over the potential barrier into the 2L material. This current depends on both band degeneracy (as in the conventional photothermoelectric effect) *and* the barrier height. Since the observed photocurrent is not dominated by a conventional photothermoelectric effect, the variations of photocurrent with applied gate bias are not fully described by variations of Seebeck coefficient with gate bias. The weak gate dependence of the photocurrent observed here may arise from carrier concentration dependent hot carrier thermalization and cooling involving carrier-carrier and carrier-phonon scattering[4], analysis of which is beyond the scope of this work.

The photovoltaic dominated photocurrent magnitude at the metal/1L junction increases with increasing gate bias due to increased band bending at the junction, as is seen in semiconducting CNT-metal junctions in accumulation[7]. This is different from a metallic CNT in accumulation where the photothermal dominated current at the contacts decreases with gate voltage.[7]